\documentclass[sn-mathphys,Numbered]{sn-jnl}

\usepackage{graphicx}%
\usepackage{multirow}%
\usepackage{amsmath,amssymb,amsfonts}%
\usepackage{amsthm}%
\usepackage{mathrsfs}%
\usepackage[title]{appendix}%
\usepackage{xcolor}%
\usepackage{textcomp}%
\usepackage{manyfoot}%
\usepackage{booktabs}%
\usepackage{algorithm}%
\usepackage{algorithmicx}%
\usepackage{algpseudocode}%
\usepackage{listings}%
\usepackage{graphicx}
\usepackage{threeparttable}
\usepackage{tablefootnote}
\usepackage{url}
\usepackage{xurl}

\theoremstyle{thmstyleone}%
\theoremstyle{thmstyletwo}%

\theoremstyle{thmstylethree}%

\begin{document}

\title[Article Title]{Cross-border offshore hydrogen trade and carbon mitigation for Europe's net zero transition}




\author[1,2]{\fnm{Sheng} \sur{Wang}}\email{sheng.wang@newcastle.ac.uk}

\author[2,3]{\fnm{Muhammad} \sur{Maladoh Bah}}\email{muhammad.maladohbah@seai.ie}


\affil*[1]{\orgdiv{School of Engineering}, \orgname{Newcastle University}, \orgaddress{\street{Newcastle University}, \city{Newcastle upon Tyne}, \postcode{NE17RU}, \country{The United Kingdom}}}

\affil[2]{\orgdiv{School of Electrical and Electronic Engineering}, \orgname{University College Dublin}, \orgaddress{\street{Belfield}, \city{Dublin}, \postcode{D04 V1W8}, \country{Ireland}}}

\affil[3]{ \orgname{Sustainable Energy Authority Of Ireland}, \orgaddress{\street{3 Park Place, Hatch Street}, \city{Dublin}, \postcode{D02 FX65}, \country{Ireland}}}



\abstract{
European countries are ambitious in both the net-zero transition and offshore energy resource development. The Irish and UK governments announced their commitments to offshore wind capacities - 37 and 125 GW, respectively, in 2050, more than two times higher than their projected power demands. While other continental countries, such as Germany, are calling for cleaner fuel resources. Exporting surplus offshore green hydrogen and bridging supply and demand could be pivotal in carbon emission mitigation for Europe. Yet, the potentials of these Island countries, are usually underestimated.
This paper developed a bottom-up method to investigate the role of offshore hydrogen from Ireland and the UK in the decarbonisation of the entire Europe. We evaluate the future hydrogen/ammonia trading and the contributions of each country in carbon emission mitigation, considering their relative cost-competitiveness in offshore hydrogen production, domestic hourly power and gas system operation, and international shipping costs.
Results indicate that the offshore green hydrogen could reduce 175.16 Mt/year of carbon dioxide emissions in Europe. The UK will be the largest hydrogen supplier from 2030 to 2040, while surpassed by Ireland in 2050, with 161 TWh of hydrogen exports to France and Spain. The offshore green hydrogen can contribute to 175.16 Mt of annual carbon dioxide emission reductions in total. This general flow of hydrogen from the West to the East not only facilitates Europe's net-zero progress, but also reshapes the energy supply structure and helps to ensure energy security across the European continent. 
}




\maketitle

\section{Introduction}\label{sec1}

Offshore wind offers a promising solution to achieving net zero ambitions in Europe, especially for island countries. For example, Ireland and the UK have recently announced their development plans for offshore wind - 37 and 125 GW installation capacities respectively in 2050 \cite{FutureFrameworkfor, Offshorewindnetzero}. These are formidably ambitious numbers because they are more than two times higher than the projection values of the overall power demands in 2050 \cite{TenYearGenerationCapacity, FES}. It is challenging to connect and utilise such large volumes of offshore wind generation with domestic power systems, simply because of the lack of demand, not to mention the real-time balancing burdens due to the uncertainties and intermittence of the offshore wind velocities.

These countries' significant offshore wind ambitions envisage an important role for green hydrogen. Green hydrogen, as produced by renewable generation through electrolysis, does not generate carbon emissions during operation compared to grey or blue hydrogen \cite{Comparisonoftheemissions}. As an energy carrier, hydrogen can be stored on a large scale more easily than electricity, as an indirect measure to store the surplus renewable generation. It can be utilised for domestic applications or generation, or transported to meet the demand at distant locations in bulk by ships or gas networks. 
While the UK and Ireland are enthusiastic about being hydrogen suppliers, some countries across the English Channel have announced their considerable need for green hydrogen in the future. For instance, Germany expects annual imported hydrogen needs of 45-77 TWh by 2030 \cite{TheNationalHydrogenStrategy, HydrogenWhichimport}, and 20 TWh for Belgium \cite{BelgianfederalHydrogenStrategy}. Linking exporting and importing countries can simultaneously ease wind connection bottlenecks and foster clean growth of the economy on a large scale, particularly for hard-to-decarbonise sectors. 

Cost is still the crucial factor that hinders the large-scale production and utilisation of offshore green hydrogen, and guides the development of the future international hydrogen trading market. The levelised cost of hydrogen (LCOH) in 2030 is projected to be around 54-184 €/MWh, which can be competitive with blue hydrogen with carbon capture and storage (CCS) and green hydrogen from onshore wind \cite{GlobalHydrogenReview2023}. In Ireland, the planned installation capacity of offshore wind is expected to first exceed the onshore wind in around 2035 due to richer wind resources, higher capacity factor, and fewer land-use restrictions \cite{windenergyroadmap}. The LCOHs in different countries and regions in Europe, including some North Sea countries \cite{rogeau2023techno}, the UK \cite{giampieri2024techno}, Poland \cite{komorowska2023evaluating}, and the Iberian Peninsula \cite{gonzalez2024techno, calado2024assessment} have been evaluated separately. They provide the projection of LCOH  from now to 2050, with detailed cost breakdowns, various hydrogen production and transmission routes, and sensitivity analysis with uncertainties. 
However, few studies have assessed the cost competitiveness of different countries in a unified framework. Yet, this is crucial for projecting the future international hydrogen trading landscape in Europe. 

Several major import routes were identified in European countries' government policies, including from North Sea countries (Denmark, Germany, and the Netherlands), North Africa, the Middle East, and America, but the potentials from Ireland and the UK, with two of the best offshore resources in Europe, are usually underestimated. 
Earlier last year, Ireland and the UK's government policies and consulting reports preliminarily identified exporting opportunities for green hydrogen \cite{HydrogenImport, FutureFrameworkfor}. However, no quantitative findings or methodologies of the offshore hydrogen cost or potential are obtained, and the ways of transportation for Ireland's green hydrogen export are limited to the pipeline to France only due to the passive cost-competitiveness estimation of LCOH. 
More importantly, studies for a single country are not sufficient for projecting the future international hydrogen trading landscape.
A synthetic analysis is urgently needed to understand how hydrogen import and export can better allocate resources, coordinately contribute to the net-zero transition of the entire Europe, as well as for the government to underpin investments and construct infrastructures ahead of time.

In this work, we for the first time investigate the roles of offshore hydrogen, especially the export from Ireland and the UK, in the decarbonization of the entire Europe. 
First, we developed a unified bottom-up framework to assess the potential and cost-competitiveness of offshore hydrogen produced by major coastal countries in Europe. High spatial-temporal resolution geo-based meteorological data and social-economic data across Europe are used to simulate the wake effect and quantity-cost supply curves of offshore hydrogen for each country.
Secondly, using Ireland as an example, we proposed a new coordinated optimisation framework for power and gas systems to evaluate the domestic integration and export potentials of offshore electricity and hydrogen. The power and gas systems are coupled by gas-fired power plants and electrolysers, where bidirectional energy exchange is allowed to further activate the flexibility of whole energy systems. Various hydrogen integration schemes (including hydrogen blending and pure hydrogen), corresponding physical constraints, and the evolution of power system operating codes (such as inertia, must-run units, and system non-synchronous generation (SNSP)) are modelled in future scenarios.
Based on the cost-supply curves and exporting potential derived above, we then formulate an optimal international trading model to project the hydrogen flows among countries in 2030, 2040, and 2050. Then, the overall landscape of offshore hydrogen in Europe's net-zero transition, as well as the contributions from each country, becomes clear. These key findings can further strengthen confidence in the continuous investment in offshore facilities and help the governments with more bespoke offshore policy design.

\section{Results}
\subsection{Cost-competitiveness of green hydrogen production in coastal European countries}

\begin{figure}
    \centering
    \includegraphics[width = 0.99\columnwidth]{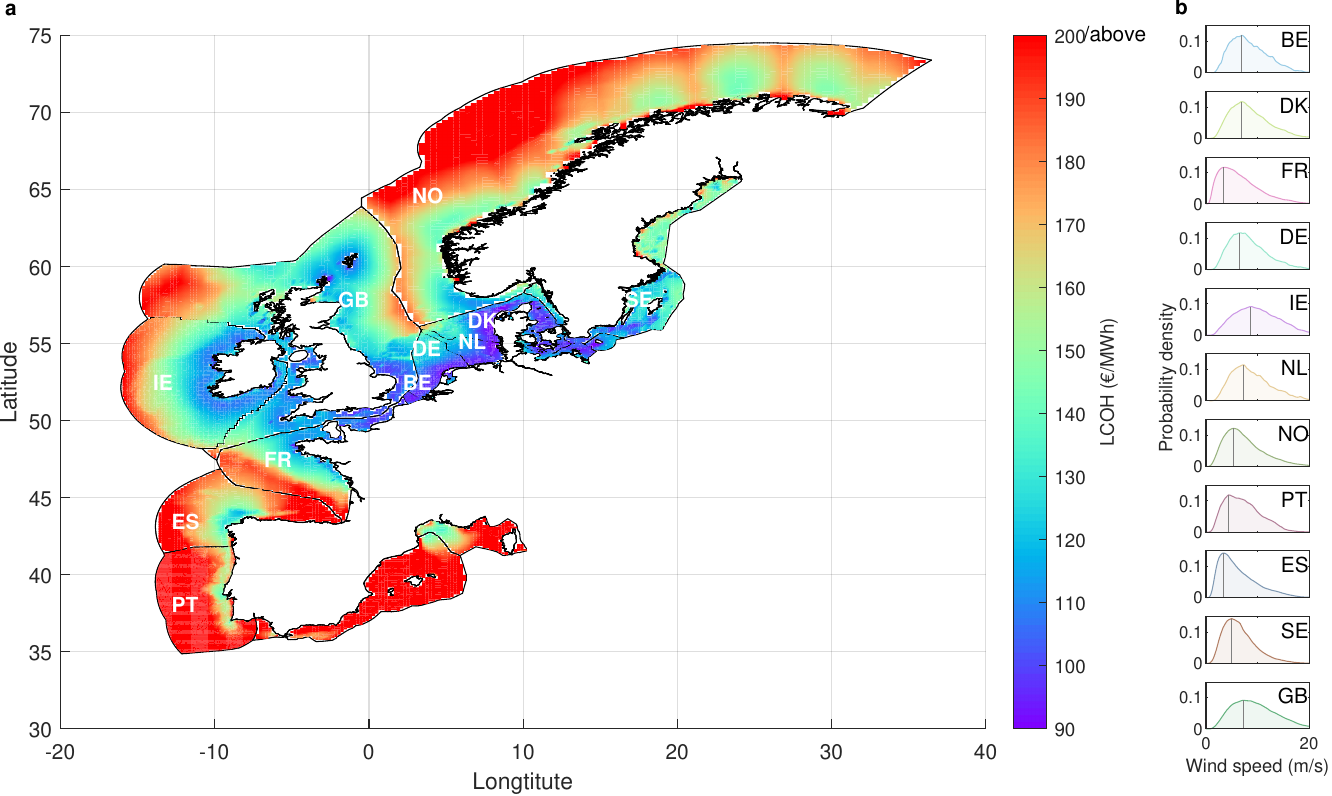}
    \caption{\textbf{LCOH map and wind speed in coastal European countries.} \textbf{a} LCOH map in EEZs of coastal European countries. The black line outlines the EEZs. \textbf{b} Probability density of wind speed for each country. The vertical grey line in each figure shows the wind speed most likely to occur. BE, DK, FR, DE, IE, NL, NO, PT, ES, SE, and GB are the abbreviations for Belgium, Denmark, France, Germany, Ireland, The Netherlands, Norway, Portugal, Spain, Sweden, and the United Kingdom, respectively.}
    \label{fig LCOH map}
\end{figure}

We first designate the major coastal countries within the assessment scope of this paper as Belgium, Denmark, France, Germany, Ireland, Netherlands, Norway, Portugal, Spain, Sweden, and the United Kingdom (UK), for these countries have large exclusive economic zones (EEZ) for offshore infrastructure development, and relatively clear and short water path to the Ireland or the UK. The historical meteorology data related to offshore wind resources modelling, such as wind velocity, air pressure, and temperatures, are obtained from the ERA-5 product at 1 hour and 0.5° x 0.5° temporal and spatial resolutions \cite{ERA5}. Hourly capacity factors, and the impact of the wake effect on power generation are simulated based on the power curve and the deployment of a 15 MW wind turbine \cite{musial2019oregon}. Probability distributions of the magnitude of the wind speed in the EEZs of the studied countries are presented in Fig. \ref{fig LCOH map}. b. It can be observed that Ireland has the richest wind resources with the highest mode (the value with the highest probability) in wind speed distribution, followed by the UK in second place, mostly owing to their access to the Atlantic Ocean. Compared to the continental countries, the probabilities of higher wind speed in these two Island countries are higher, with richer and more evenly distributed wind resources across the whole year.

Based on the capacity factors, taking account of the development, capital, operational, and decommission expenditures of the offshore hydrogen production platform (including wind turbine, foundation, electrolyser, etc., as the method can be found in the supporting information), the LCOH map can be created in Fig.\ref{fig LCOH map}. a. Bathymetric data, port and connection point locations to the power and gas networks are considered in the model, so that the impacts of water depth on materials and the distance to the maintenance costs can be quantified \cite{bathymetry_2022, EMODnet}. We can find that the LCOH ranges from 84.93 to over 200 €/MWh in EEZs. The lowest LCOH appears in Denmark due to its high-level capacity factors and shallow continental shelf. This price becomes competitive, but is still 33.94 \% higher than onshore blue hydrogen 68.92 €/MWh \cite{Globalaveragelevelised}. The UK's LCOH ranges from 85.97 to 307.26 €/MWh, with the second lowest lower bounds for LCOH. The LCOH of Ireland ranges from 90.42 to 201.8 €/MWh, which is not as highly competitive as compared to the leading countries. 

\begin{figure}
    \centering
    \includegraphics[width = 0.99\columnwidth]{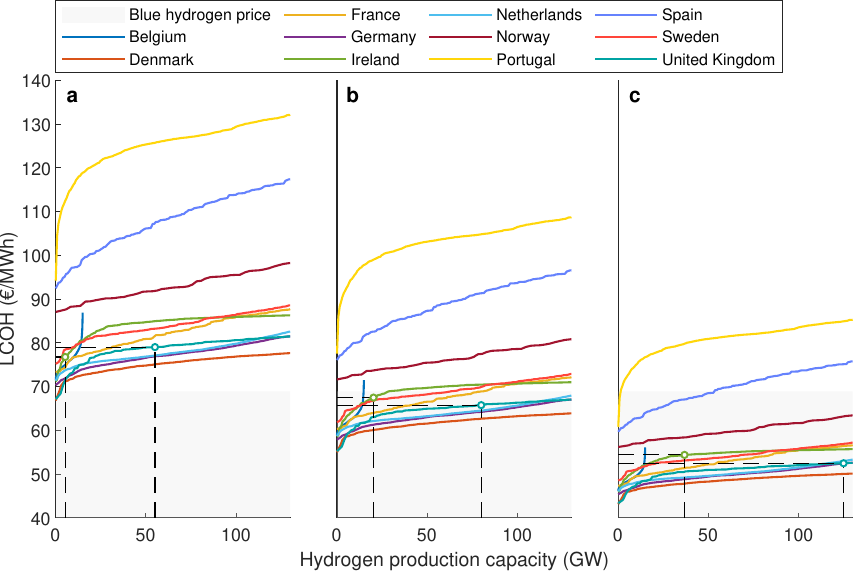}
    \caption{\textbf{Hydrogen cost-supply curves for different countries at different years.} 
    The cost-supply curves of green hydrogen produced by countries in 2030, 2040, and 2050 are shown in \textbf{a}, \textbf{b}, and \textbf{c}, respectively. Different colours of solid lines represent different countries. Green and blue dots on Ireland's and the UK's curves marked their goals for offshore wind capacities. Grey areas represent the possible range of blue hydrogen production prices in the future.}
    \label{fig LCOH curve}
\end{figure}

\begin{figure}
    \centering
    \includegraphics[width = 0.8\columnwidth]{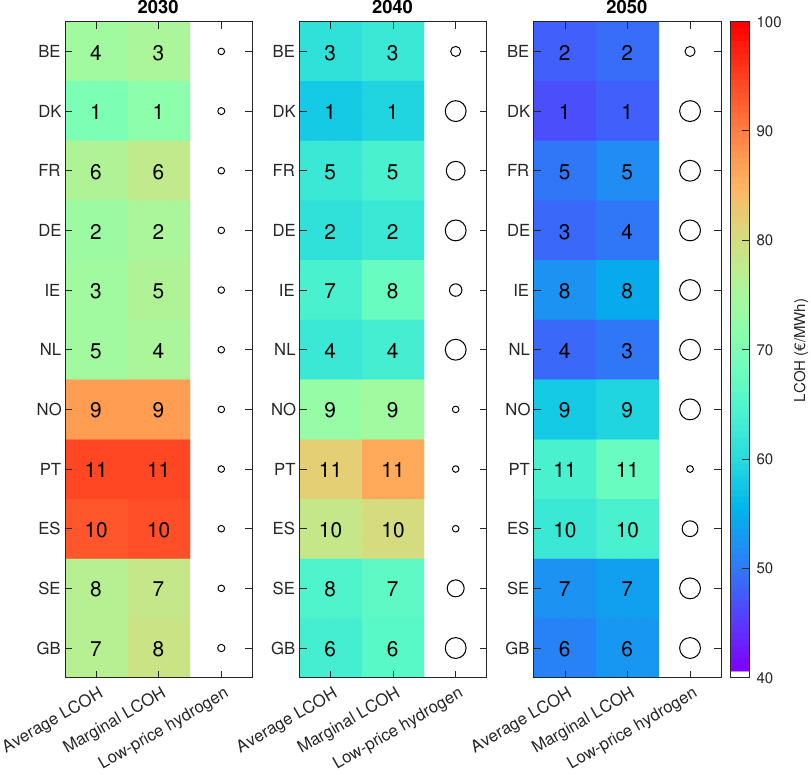}
    \caption{\textbf{Cost-competitiveness of offshore green hydrogen among major countries and with blue hydrogen.} 
    Rank means the ranks of average/marginal LCOH among 11 countries; Marginal means the LCOH at the offshore capacity of that country in the corresponding year; Average means the LCOH of sum-up average of all offshore wind capacities in the corresponding year; Low-price hydrogen means the capacity of hydrogen production whose production cost is lower than the upper bound of blue hydrogen. The size of the circle indicates the capacity of low-priced hydrogen. The smallest and largest size indicates 0 and >100 GW, and other sizes indicate the capacities between these two values.}
    \label{fig lcoh rank}
\end{figure}

Further considering the influence of the wake effect, shipping density, and excluded areas (such as military areas, and aquaculture sites) on the power density when deploying offshore wind turbines, and the cost reduction in the future \cite{wiser2021expert}, the cost-supply curves of hydrogen for 2030, 2040, and 2050 scenarios are presented in Fig. \ref{fig LCOH curve}. 
In 2030, the marginal cost of hydrogen in Ireland and the UK are 76.10 and 79.02 €/MWh, respectively, as shown in the Table. \ref{fig lcoh rank}, with Ireland's a bit lower than the UK's. However, if they are producing the same volume of hydrogen, Ireland's LCOH is generally higher than the UK's. Ireland has a competitive advantage over  Portugal, Spain, Norway, and Sweden initially, but its LCOH exceeds Sweden's after the capacity reaches 10 GW. Within 5 GW, the UK has a similar competing strength to Denmark in the first place, which ensures the UK's smooth start for the offshore hydrogen market. After 5 GW, the UK still has consistent advantages over most of the other countries. However, due to the more ambitious offshore capacity, the UK's marginal LCOH is ranked in 8th place, which means other countries with higher ranks, such as France and the Netherlands, will still have a certain volume of competitive hydrogen supplies regardless of their higher LCOH cost-supply curves. At this stage, most countries' offshore green hydrogen production cannot compete with blue hydrogen. 

In 2040, owing to the advancements of technologies and supply chains, the LCOH will reduce by around 21\% \cite{wiser2021expert}. With the increase in offshore goals, their cost competitiveness also changes. Ireland's marginal LCOH is 67.40 €/MWh, beginning to exceed the UK, Sweden, and France, due to its ambitious offshore goals. The UK's rank has moved forward and displaces Sweden, owing to its continuous advantageous endowment to the offshore wind (capacity factor, water depth, and long coast). Denmark is still in the leading position. At this stage, more countries can produce offshore green hydrogen, which has the chance to compete with blue hydrogen, especially Denmark, the Netherlands, and the UK.
As the time approaches 2050, the marginal LCOHs of Ireland and the UK further reduce to 52.51 and 54.41 €/MWh, respectively. While the rank of the UK remains the same, Ireland becomes less competitive in terms of both average and marginal LCOHs, due to its ambitious offshore goal and less favourable offshore conditions. At this stage, with the further reduction of LCOH and relatively constant blue hydrogen price due to the increase of fossil fuel and carbon costs, all these countries can begin to have the opportunity to compete with blue hydrogen. 

In summary, Denmark is the most cost-competitive country, which has the overwhelmingly lowest LCOH from 0 to 150 GW capacity. Ireland and the UK generally lie in the middle, and may have a certain cost competitiveness depending on the supply-demand relationship in the hydrogen market (the analysis can be found in the following sections). From now to 2050, because of the relatively steady blue hydrogen price \cite{Globalaveragelevelised}, offshore green hydrogen will finally become cost-competitive in all these European countries. 

\subsection{Domestic offshore wind integration and hydrogen export potential} 

Because the ambitious offshore wind capacities in Ireland and the UK are much higher than their power demands, we analyse the integration potential of domestic energy systems to determine the export potential of green hydrogen. A whole energy system optimisation approach is designed to simulate the operating conditions hour by hour across a whole year with uncertain offshore wind generation. The power and natural gas systems are deeply interconnected through gas-fired power plants and on/offshore electrolysers, which are optimised simultaneously to expand the flexibility for the accommodation of offshore wind fluctuations. Green hydrogen can be blended into the natural gas network at a variant ratio, so it can be transported in bulk, and the analysis can cover various future scenarios from zero hydrogen to pure hydrogen networks. 
Hourly unit commitment is conducted considering all available generations in the All-Ireland power system (including coal, waste, peat, natural gas, oil, onshore wind, solar, hydro, and offshore) and practical operating codes designed by EirGrid \cite{OperationalPolicy} for future operation scenarios (such as ramping, interconnectors, inertial constraints, must-run units, and SNSP), actual hourly power demand curves are accounted (see the supporting information for detailed methodologies). 

\begin{figure}
    \centering
    \includegraphics[width = 0.99\columnwidth]{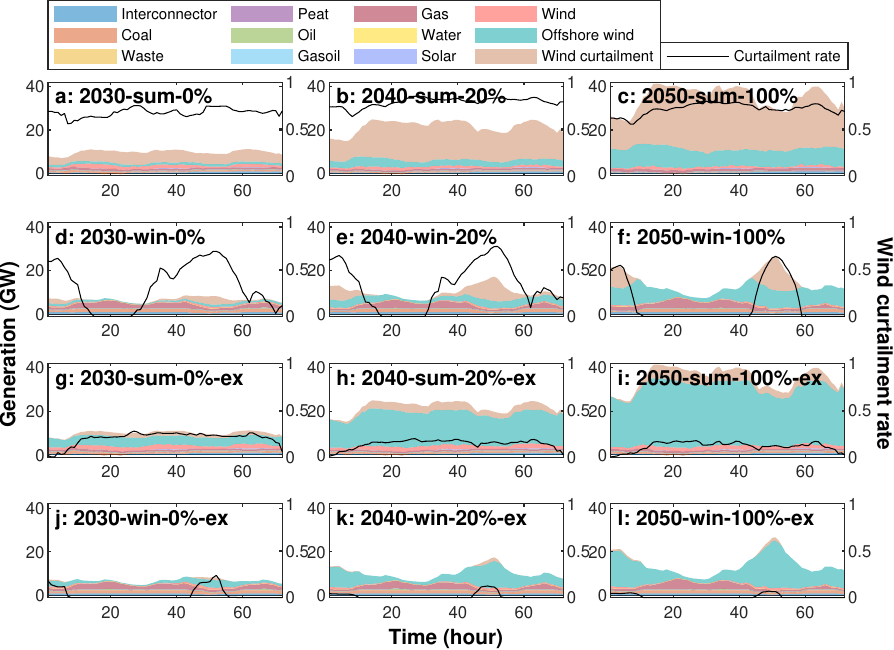}
    \caption{\textbf{Power system operation results.} Different years, seasons, hydrogen blending ratios, and export conditions, are compared. 2030, 2040, and 2050 indicate the year; sum and win mean a typical summer day (high wind and low demand) and a typical winter day (low wind and high demand), respectively; 0\%, 20\%, and 100\% mean different maximum allowed hydrogen blending potential; ex means export.
   }
    \label{fig unit commitment}
\end{figure}

The unit commitment results and wind curtail rate in different seasons, years, and with/without exporting options are compared in Fig.\ref{fig unit commitment}. Considering that the UK government has allowed 20\% hydrogen blending at the distribution level in specific areas, it is assumed the gas transmission network is ready for a maximum 20\% blending ratio in 2040. In 2050, pure hydrogen networks are constructed in line with the UK's hydrogen network roadmap \cite{BritainsHydrogenNetworkPlan}.
As we can see from Fig.\ref{fig unit commitment} a to Fig.\ref{fig unit commitment} c, from 2030 to 2050 in summer, with the increase in offshore wind capacity, the average wind curtailment rate without export generally increases from 67.51\% to 71.29\%. The high wind curtailment is mainly caused by the large difference in high available wind generation and low demand.
In the winter, owing to the fewer wind resources, the average wind curtailment rate decreases by 56.25\% from 2030 to 2050. This is because, in this scenario, the increase in hydrogen blending ratio plays a dominant role, which allows the gas system to absorb more wind. 

However, we can observe that even after replacing all natural gas with hydrogen and fully exploiting the flexibility of the energy system, a considerable proportion of offshore wind will still be curtailed mainly due to low domestic energy demand. This highlights the necessity of exporting hydrogen to other countries. 
The hydrogen export access could significantly promote offshore wind utilization across all years and all seasons of scenarios, if we compare the upper six figures (Fig.\ref{fig unit commitment} a-f) with the lower six figures (Fig.\ref{fig unit commitment} g-l). For example, the average wind curtailment rate in the 2050 summer can be reduced to 11.53\%. The exporting potential for Ireland is huge (average 27 GW in Fig. \ref{fig unit commitment} i, even 1.68 times higher than total domestic energy demand).
The uncertainty of offshore wind requires operating reserves to be scheduled ahead of real-time operation. High costs of operating reserves and policies on reserve capacity, inertia, and must-run units prevent energy systems from accommodating a certain portion of marginal offshore wind. However, this marginal value does not grow with the year and offshore capacity, so it is interesting to observe that with export options available, the wind curtailment rate even decreases with the increase of offshore wind capacity, which is opposite to the case without export options.

\begin{figure}
    \centering
    \includegraphics[width = 0.99\columnwidth]{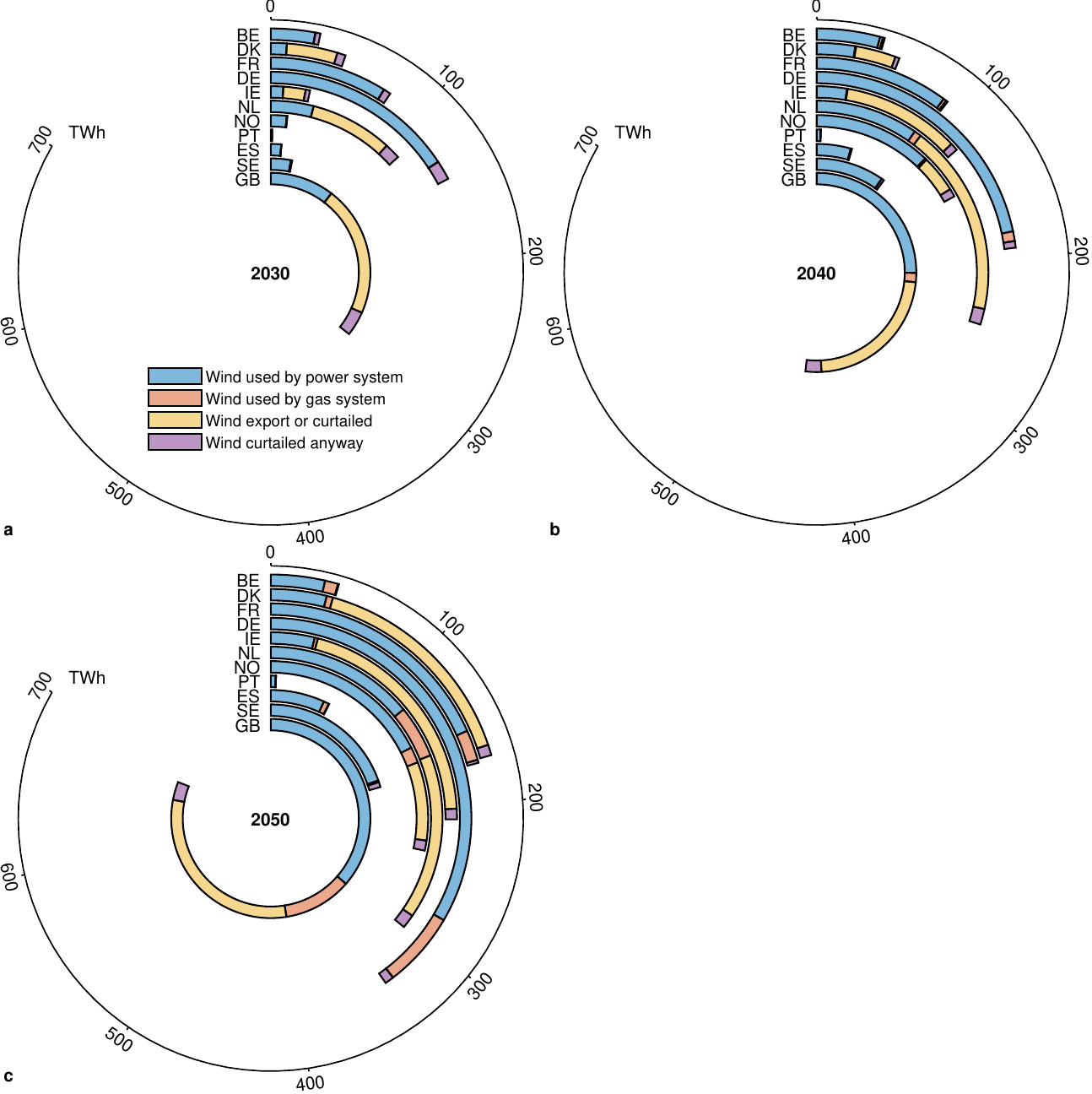}
    \caption{\textbf{Utilisation of offshore wind.} 
    \textbf{a}, \textbf{b}, and \textbf{c} decouple the utilisation of wind in 2030, 2040, and 2050, respectively.}
    \label{fig wind decomposition}
\end{figure}

Using the whole year's results from Ireland, we estimated the offshore wind utilization status for other countries. We assume the offshore wind consumption is proportional to the electricity and gas demand.
The annual offshore wind utilization is decomposed in Fig. \ref{fig wind decomposition}. Among these countries, the UK is leading in offshore wind capacities, and the domestic load is not very high, thus it has the highest exporting potential of 178 TWh in 2030. The Netherlands, Denmark, and Ireland are following after it, with 65, 30, and 16 TWh exporting potential, respectively. 
In 2040, although the UK's offshore capacity growth is relatively slower than its energy demand, it and the Netherlands are still the top two exporting countries. On the other hand, because Ireland's offshore capacity has quadrupled to 20 GW while the energy demand increase is much slower, its exporting potential reaches 86 TWh, surpassing Denmark and becoming the third-largest hydrogen exporting country.  
In 2050, with access to 100\% hydrogen in natural gas systems, the offshore wind consumption potential from domestic energy systems will increase, especially for countries relying on natural gas generation. Therefore, although the offshore capacities further increase, the exporting potential of the Netherlands even decreases by 13.87\%. 
For countries with relatively lower offshore potential and higher domestic energy demands like Germany and France, although their offshore wind capacities are also considerable, there is no green hydrogen left for export. 
On the contrary, the hydrogen export potentials from the UK and Ireland continue to increase significantly by 32.59\% and 96.82\% respectively compared with 2040. The UK is leading first place with 232 TWh capacity. Notably, Ireland, with 167 TWh exporting availability, due to its large offshore capacity and low energy demands, exceeds Denmark and the Netherlands to become the second-largest hydrogen exporting country in Europe.

\subsection{Carbon emission mitigation potential in Europe}

\begin{figure}
    \centering
    \includegraphics[width = 0.7\columnwidth]{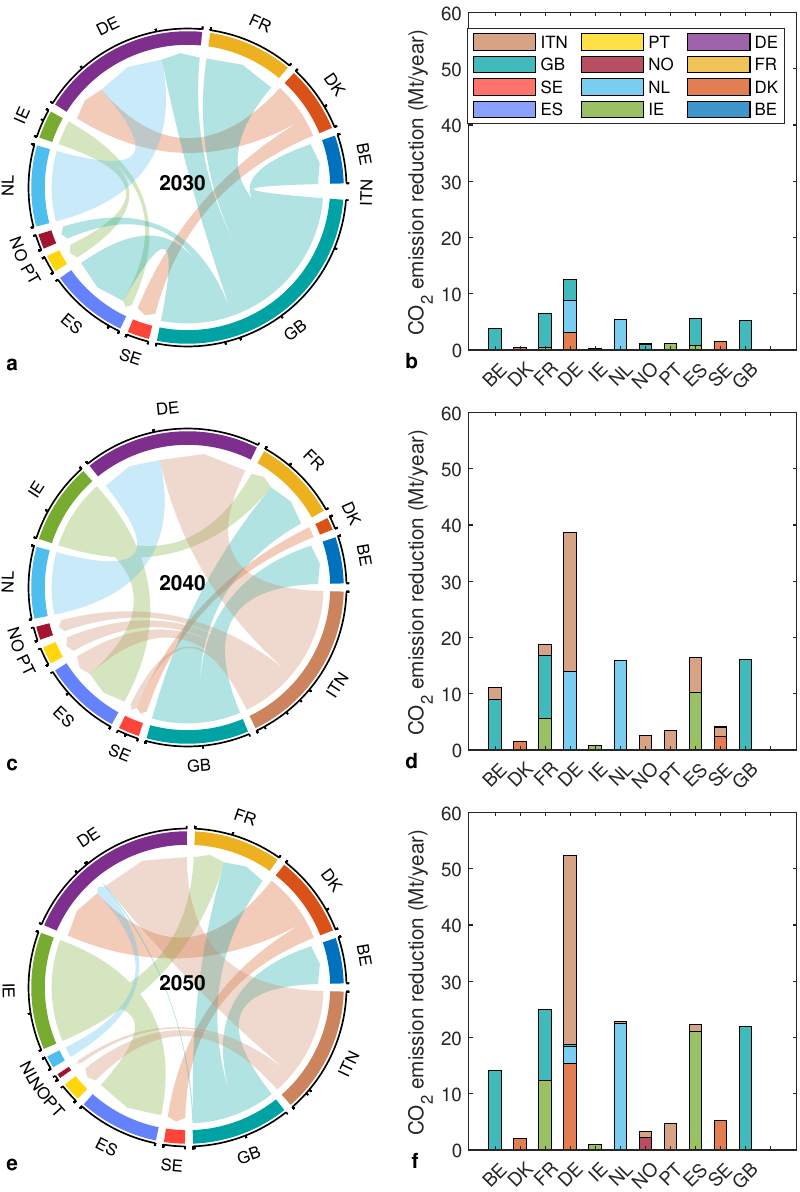}
    \caption{\textbf{Sankey diagram of hydrogen flows and contributions to decarbonization} 
    \textbf{a}, \textbf{c}, and \textbf{e}, are the Sankey diagrams of hydrogen flows among these countries and international hydrogen imports in 2030, 2040, and 2050, respectively. ITN represents international hydrogen import. Please note that only the hydrogen flows between countries are drawn. The hydrogen supplied by and to itself is not indicated in these figures. \textbf{b}, \textbf{d}, and \textbf{f} show the contributions to the decarbonization of each country by international green hydrogen trading. Different bars represent the reduction in carbon dioxide emissions, and different colours indicate different countries' contributions.}
    \label{fig sankey}
\end{figure}


By combining the cost-supply curves and exporting potential analysis in previous sections, here we developed an optimal international green hydrogen trading model to simulate the hydrogen flows among countries in future scenarios. The cost-supply curves are interpolated as piecewise linear functions. Together with shipping costs, they are modelled as objective functions and the optimisation model maximises the decarbonisation with minimal cost. The Sankey diagram of hydrogen flows, reduction of carbon emission, and future hydrogen trading landscape, are shown in Fig. \ref{fig sankey}.

In 2030, the UK will be the major hydrogen exporter, followed by the Netherlands, Denmark, and Ireland. It contributes to 98.29\%, 98.38\%, 26.08\%, and 83.99\% of the hydrogen demands of Belgium, France, Germany, and Spain, respectively. 
Germany is the major hydrogen receiver due to its high demand. Besides the UK, 25.72\% and 46.80\% of Germany's hydrogen demand vacancy is fulfilled by Denmark and the Netherlands, owing to the low hydrogen production cost, nearby locations and consequently lower shipping cost. 
When it comes to 2040, a major change in the hydrogen trading structure is that the international hydrogen import becomes essential in fulfilling the hydrogen demand, due to the slow offshore capacity growth relative to the hydrogen demand growth. The share of the UK in the hydrogen supply market decreases dramatically. It is no longer able to supply Germany's hydrogen demands after meeting Belgium and France. As a result, Germany will have to import 102 TWh of hydrogen from international markets (such as Eastern Europe and South Africa). The hydrogen demands in Norway, Portugal, and Spain are partially covered by international imports as well. In contrast, we can witness the growth of Ireland's market share at this stage, which supplies 28 and 52 TWh of hydrogen to France and Spain, respectively.
In 2050, as the electrification approaches the end phase and the increase in hydrogen demand slows down, the hydrogen import from the international market will reduce, but will still play an important role in the overall hydrogen supply structure, and will continue its contribution to Germany and Portugal. The reduced share of international hydrogen and the Netherlands is now covered by Ireland. 
Though the hydrogen production capacity of Ireland is still lower than the UK, the hydrogen export flow increases to 161 TWh, for the first time exceeding the UK and becoming the largest hydrogen exporter in Europe. 

The contributions of hydrogen export and import are also reflected in the reduction of carbon emission, as shown in Fig. \ref{fig sankey} b, d, and f. In general, the offshore hydrogen production from these 11 countries can contribute to 43.53, 129.32, and 175.16 Mt annual carbon dioxide emission reductions in 2030, 2040, and 2050, respectively, equivalent to 0.74, 2.20 and 2.98 times of Ireland's current carbon emission \cite{COemissions}. In 2030, the UK and the Netherlands will be the main contributors to carbon reduction. Then from 2040 and 2050, similar to the hydrogen supply structure, the importance of Ireland increases gradually. Ireland and the UK, not only account for all carbon mitigation in their own country (1.05 and 21.93 Mt, respectively), but also help the decarbonization of other countries in Europe (33.45 and 27.21 Mt in total).

\section{Discussions}

The overall analysis conducted in this paper shows great potential for offshore green hydrogen on the entire Europe's net zero transitions, especially the export potential from two Western European Island countries, Ireland and the UK.
Denmark has a considerable competitive strength in offshore hydrogen supply from now to 2050, while Ireland and the UK are in a middle position due to their rich wind resources and rich bathymetry endowment. 
With the reduction of technology costs in the long term, the LCOH can be reduced to 43.15 €/MWh. Over 900 GW of offshore green hydrogen capacity will have the chance to compete with blue hydrogen, owing to the relatively steady fossil fuel price and CCS infrastructure cost. 
The domestic offshore wind and hydrogen integration analysis shows that Ireland and the UK not only have strengths in cost, but also demonstrate considerable exporting potential. Ireland will surprisingly have the second-largest available hydrogen exporting capacity of 167 TWh in Europe in 2050, due to its ambitious offshore policies and low domestic energy demand. Other countries like Germany have higher offshore capacity, but the even higher domestic energy demand consumes all offshore renewable energy and consequently leaves no hydrogen exporting capacity. In return, hydrogen export capacity can also be considered a flexible resource, which can help to improve the wind utilization rate and mitigate the offshore wind connection burden for Ireland's domestic power and gas systems.
Finally, our international hydrogen trading model linked the cost-competitiveness and export potential models to predict the future landscape for cross-border green hydrogen trading. Through hydrogen trading, the carbon dioxide emissions in Europe can be directly reduced by 175.16 Mt/year in 2050. 
From 2030 to 2050, Ireland and the UK will grow into the two largest green hydrogen suppliers in Europe. They account for 399 TWh of annual green hydrogen supply capacity in 2050, contributing to 47.75\% carbon emission reduction for Europe's net zero transition.
 
Despite the Irish government having expressed a general favour for offshore hydrogen production and export, it remains cautious and unclear about the competitiveness against other European producers.
In 2024, Ireland published the future framework for offshore renewable energy policy statement along with supporting documents from consulting companies \cite{OffshoreRenewablesSurplus}, and consider Ireland as broadly competitive globally but maintains no outright competitive advantage relative to other European production economies. Furthermore, the reports suggest that a considerable volume of surplus economic hydrogen would be available after domestic demand is met, which can be utilised for exports. Therefore, the Irish government is inclined to explore pipeline transportation options (e.g., building a pipeline to France from repurposed gas pipelines) to reduce the transportation cost and to be economically advantageous. However, if hydrogen export is limited to pipeline transportation, the capacity and flexibility will be constrained.
However, we would like to highlight that our research indicates promising alternatives. 
Ireland has no strong competing advantage in LCOH, and does not have the largest offshore hydrogen production capacity, but it will still become the largest contributor to the decarbonisation of the entire Europe through green hydrogen export. This is because of a combination of reasons, including fair competitiveness, large export potential, and general hydrogen insufficiencies in Europe. 
This eye-refreshing conclusion can only be derived through our bottom-up optimisation approach, which has never been reported before in previous research or government policies. By combining the isolated cost-supply curves with other holistic energy systems and international trading optimization models, we have identified more opportunities for Ireland in the future hydrogen market.

Finally, we would like to highlight that our results on the contributions of offshore hydrogen to the decarbonization of Europe are calculated on the basis of currently published offshore goals, with the assumption that these infrastructures will be delivered on schedule. 
Our analysis indicates that there is still a need for international hydrogen imports from other areas of the world (such as Eastern Europe, the Middle East and North Africa) with significantly higher costs or unstable supply conditions to fulfil the shortage of hydrogen demand in 2040 and 2050. This means, despite the very ambitious offshore goals we already have, there is always space to do more.
First, more ambitious offshore hydrogen production goals can be set to cover the 162 TWh international hydrogen import in 2050, equivalent to about 37 GW of offshore electrolysers and corresponding offshore wind farms. Based on our analysis in Section 2.1, all these countries except Belgium still have the potential to further increase their offshore capacities. This is especially recommended for countries with lower marginal costs, such as Denmark and the Netherlands. 
Second, we could accelerate to address the non-technical bottlenecks in offshore infrastructure construction, such as supporting domestic and international investments (from Denmark, US, or China) in industrial manufacturing capacity in offshore wind turbines, electrolysers, subsea cables, and hydrogen shipping vessels. The government will need to provide clear and stable policy frameworks to de-risk private investment, accelerate consenting processes for offshore wind farms, and establish dedicated hydrogen export corridors. Then, we can reduce the temporary reliance on international hydrogen imports in 2040, and guarantee a smoother expansion of the hydrogen market. 

These quantitative discoveries could help European governments revisit offshore goals when they constantly update their hydrogen national strategies. Our findings visualised a very promising hydrogen market potential and continuous shortages in offshore infrastructure investment, which could indicate attracting more investments. 
This west-to-east offshore hydrogen exporting from will significantly reshape the energy flow pattern in Europe.  
It can not only make wider contributions to the net-zero transition, but also enhance the national energy security of European continental countries, which heavily rely on international gas imports (e.g., from Russia), and ultimately solve the energy trilemma.

\section{Methods}\label{sec11} 

\subsection{Tech-economic model of offshore hydrogen production}

The offshore wind generation and cost are assessed within the EEZs of the selected coastal European countries, i.e., Belgium, Denmark, France, Germany, Ireland, Netherlands, Norway, Portugal, Spain, Sweden, and the UK \cite{Marineregions}. Aquaculture Sites, Special Areas of Conservation, and military areas were excluded from the available offshore hydrogen production areas, the geo-boundary of which is obtained from the European Marine Observation and Data Network (EMODnet) \cite{EMODnet}.
Within these geographical boundaries, vectorized wind velocity data, as well as other weather data that could affect wind generation (temperature and air pressure) are obtained from the ERA-5 product of the fifth-generation European Centre for Medium-Range Weather Forecasts (ECMWF) reanalysis for the global climate and weather in the last five years. Data are provided at 1 hour and 0.5° x 0.5° temporal and spatial resolutions (equivalent to about 45km by 56 km in South Europe) for the last five years \cite{ERA5}. 

Wind power generation and capacity factors are calculated on an hourly basis using the power curve of an NREL reference 15 MW wind turbine, where the technical parameters are described in \cite{musial2019oregon}. The power curve is described at 1m/s wind speed resolution. For fractional wind speed, linear interpolation is used to estimate the power of the wind turbine.
Turbine layouts are optimised by balancing the power density and interference from wake effects. The wake effect of a single turbine is simulated using the Jensen-Gaussian model, where the effectiveness has been demonstrated in \cite{tao2019optimal}. 
When characterising the cumulative wake effects from multiple upstream-located wind turbines, multiple wake effects could overlap. Previous studies usually consider a single layer of wake effect only or use enumeration-based methods, which will increase the computation burden exponentially with the number of mutually interfered wind turbines and massive historical data \cite{paul2019multi, tao2020joint}. To overcome this challenge, a recursive-tree-based method is developed, which in practice could reduce the simulation time by 79\% with less than 1\% sacrifice of accuracy \cite{Figshare}. 
The specific methodologies and parameter values can be seen in the supplementary information.

The LCOH of the offshore wind farm is calculated using a bottom-up approach, including four cost modules, development expenditure (DEVEX), capital expenditure (CAPEX), operational expenditure (OPEX), and decommissioning expenditure (DECEX). These cost modules are further broken down into different terms and components, as shown in Fig. x in the supplementary information. Bespoke tech-economic parameters for Europe are obtained from Offshore Renewable Energy Catapult, including environment survey cost, wind turbine cost, cable cost, etc \cite{OffshoreWindandHydrogen}. 
Generally, given all the parameters of the model, it requires five additional inputs, i.e., water depth, capacity factor (as calculated above), distance to the nearest port and connecting point, and distance between wind turbines to generate the geolocation-based LCOH. 
Water depth (Bathmetry data) within the EEZs of the studied countries is obtained from the Bathymetry database published by EMODnet \cite{bathymetry_2022}. Data were converted and applied to grid cells with 0.001° $\times$ 0.001° temporal and spatial resolutions (equivalent to about 0.11 km by 0.06 km near the UK). 
The water depth affects the type of wind farms as well as the amount of material used for building the platform. For water depths less than 30m and between 30-60m, the potential of monopile structure and jacket structure (fixed-bottoms) wind turbines is explored. The deployment of floating turbines in deeper water environments (over 60m) is also evaluated, but would entail significantly higher costs both for installation and maintenance. For EEZs of studied countries, the North Sea areas are relatively shallow and can be deployed with fixed bottom wind turbines, while the water depths of Iberian Peninsula marine areas increase faster with the distance to the coast. 
The minimum distance to the port affects the installation cost when shipping materials to offshore locations, as well as the annual maintenance expenses. The major port and city locations in the studied countries are obtained from EMODnet \cite{EMODnet}.
The minimum distance to connecting points determines the capital cost of export cables that transmit electricity generation from offshore to onshore substations.
Distances between wind turbines are determined based on the principle that the maximum interference from adjacent wind turbines from the wake effect should be less than 5\%. It affects the cost of array cables that collect the generations from individual wind turbines and converge on the offshore substation.

In the offshore electrolysis platform, there are several components consuming the energy from offshore wind, including water pumps and desalination, electrolysers, hydrogen compressors, and standby batteries. 
The techno-economic characteristics of the Alkaline Electrolyser (AEC) are summarized in Table 3 in the supplementary information. The AEC system is currently the most mature and durable among these three options, and therefore is studied in this paper. The data on the capital cost and electrical efficiency were obtained from projections by the Offshore Renewable Energy Catapult and the UK's National Hydrogen Strategy. The efficiencies of AEC are projected to increase from 0.79 in 2030 to 0.82 in 2050 \cite{HydrogenAnalyticalAnnex}.
The capital cost includes previously mentioned key components, as well as necessary auxiliary systems such as platforms and pipelines. Annual OPEX includes maintenance, stack replacement, and labour costs are expressed as a percentage of initial capital investments. 
Due to the fluctuations of offshore wind and mismatch with the electrolyser's operating point, the electrical and hydrogen storage facilities are essential. Water pumps, desalination, and hydrogen compressors should be sized based on hydrogen production. Therefore, their total power consumption is dependent on the hydrogen production capacity and adds up to the rated power of the wind generation with a certain margin. On this basis, a set of algebra equations (please see Section 3.3 in supplementary information) is formulated based on the maximization of utilisation rate to determine the optimal size of the electrolyser and other accessories. The operating technical parameters and various costs in the future are projected by \cite{OffshoreWindandHydrogen}. By accounting for the DEVEX, CAPEX, OPEX, and DECEX of these devices, the LCOH can be calculated.

In addition to the LCOH map, to obtain the cost-supply curves of hydrogen, the available supplied quantity of hydrogen at each level of LCOH should be further evaluated. The potential installed
offshore wind capacity for each grid cell was calculated by multiplying the turbine's rated power by the number of turbines that could fit maximally into an individual cell. Distances between wind turbines are determined as aforementioned to control the wake effect below the threshold while maximising the power density. The vessel density could affect the power density given a specific layout of a wind farm, which is obtained from EMODnet with 1 km resolution \cite{RouteDensityMap}. It is assumed that he power density could decrease linearly with the increase of vessel density up to 50\% \cite{guo2023grid}.
In this study, we consider three future scenarios for LCOH reductions (high, median, and low) based on expert elicitation surveys \cite{wiser2021expert}. Compared with 2020, the reduction rates in 2030, 2040, and 2050 in the example of median scenarios are projected at 21\%, 35\%, and 49\% for fixed wind farms, and 2\%, 17\%, and 40\% for floating wind farms, respectively.

\subsection{Domestic offshore wind and hydrogen integration model}

The offshore power generation is integrated into the domestic power system via marine cables, while for bulk utilisation, the hydrogen production is connected to the domestic national natural gas system via pipelines through hydrogen blending techniques (assumed from 0\% to 100\% from now to 2050). Hydrogen blending refers to injecting hydrogen into existing natural gas pipelines, being fully mixed, and transported to other locations for further use. Compared with local hydrogen utilisation or dedicated hydrogen pipelines, hydrogen blending can provide a cost-efficient interim solution to rapidly promote the scale of hydrogen demand. Hydrogen blending has received substantial support and has been successfully demonstrated in numerous countries, including the European hydrogen backbone \cite{EuropeanHydrogenBackbone}, HyDeploy in the UK \cite{HyDeploy}, and HyBlend in the US \cite{HyBlend}. The UK's Department for Energy Security and Net Zero permitted distribution-level hydrogen blending in December 2023, and is actively seeking more evidence to extend it to transmission level \cite{HydrogenblendinginGB}. According to the Energy Network Association, the UK aims to achieve 100\% hydrogen networks by 2050 \cite{BritainsHydrogen}. 
Therefore, the utilisation and integration of offshore power and hydrogen production is constrained by the national energy systems, in terms of demands, network transmission capacities, and operation schemes. To evaluate the potential of the domestic energy system for accommodating offshore resources, especially for fully exploring island countries like Ireland with abundant offshore resources and inadequate domestic energy demands, a coordinated optimal operation model (COM) for the national energy system, including both power and gas networks, is developed. The optimisation model will be conducted on an hourly basis throughout the year, simulating the optimal system operating state with the lowest operation cost while maximising the offshore resource integration.

The power system module in the COM constitutes a unit commitment framework to schedule the resources in the day ahead for balancing the variation of energy demands and uncertain renewable generation on the nodal scale. The unit commitment cost includes fuel costs and start-up costs for thermal units. For the All Ireland Island power system, the practical operational policy roadmap from 2023-2030, as outlined by EirGrid and Soni, is incorporated in the model, such as Inertia constraints, Minimum Number of Conventional Units, System Non-Synchronous Penetration (SNSP) \cite{OperationalPolicyRoadmap}.
The flexibility constraints for generators are incorporated, including unit ramping limits, minimum online-offline times, minimum output levels and must-run constraints. 
The power network congestion, which may also hinder the integration of offshore wind generation, is modelled using the DC power flow model. The dynamic rating of power lines with temperature variations is considered in summer and winter scenarios. 
Future construction plans for interconnections are considered, such as the 700 MW Celtic Interconnector with France in 2027 \cite{CelticInterconnector}.

The natural gas system module is constructed to minimise the operating cost for gas systems while accommodating the hydrogen injection from offshore hydrogen. With the stochastic hydrogen injection, the gas compositions and corresponding physical properties (such as heat value and specific gravity) of the resulting gas mixtures will be uncertain, which can bring threats to the gas security and challenges to the optimal operation of gas systems. To overcome these issues, we first defined a gas security region according to the recently updated The Gas Safety (Management) (Amendment) Regulations 2023, as proved by the UK's Legislation \cite{TheGasSafety2023}, which uses relative density, Wobbe index, and additionally Weaver flame speed factor, to limit the hydrogen fraction. The weymouth equation is used to describe the relationship between gas flow rate and pressure drops along a gas pipeline to avoid over-/under pressure issues in the network. However, due to the variant gas physical properties, the Weymouth equation becomes more nonconvex, which could lead to local optima and unstable solutions. Here we developed a reformulation method and solution algorithm, comprised of bespoke McCormick envelope, second-order cone reformulation, and sequential convex programming. Then, the optimisation problem can be robustly and efficiently solved by off-the-shelf commercial solvers (such as Gurobi). Compared with general nonlinear solvers (such as IPOPT), the computation efficiency can be improved by over 90\% \cite{wang2023optimal}.
When multiple pipelines intersect, different compositions of gas will be mixed evenly at the joint point. This process is also incorporated in our model. Identically, this process will generate bilinear constraints and cause computation burdens, which is also addressed by our previously introduced solution framework (mathematical models can be found in Section 4 of Supporting Information).

For an accurate description of power and gas system operations, high temporal-spatial precision and long-span power generation, consumption, and transmission datasets are employed. On the generation side, the detailed operational parameters of All-Ireland Island generators, such as capacity, and piecewise heat rate curves, are obtained from the Single-Electricity-Market (SEM) Committee, which has been used and validated in EirGrid's publications \cite{SEMPLEXOS}. For non-gas fossil fuel units such as coal and oil, the commodity cost and heat value are also counted based on the SEM model. For gas-fired units, the fuel cost is counted on the gas system side, and the heat value is modelled dynamically with the variation of hydrogen fractions. For onshore wind and solar generations, the wind speed and solar radiation are obtained from ERA-5, and the Beckman model is used to calculate the output of solar generations. The hydro generation data are obtained from historical operations \cite{SystemandRenewableDataReports}. 
For the consumption sector, the trends of gas demand in the future scenarios are predicted according to the Gas Forecast statement by Gas Network Ireland \cite{GasForecastStatement2022}, with various cross-sector considerations,
such as the decommissioning of coal-fired power plants (e.g.,
Moneypoint units), 80\% renewable generation ambition from
Irish government’s Climate Action Plan, and impacts of
banned new gas boiler connections and Gross Domestic Product
(GDP) on residential, industrial and commercial demands,
construction of new compressed natural gas stations. As a result, the gas demand will increase in the short term,
reach a peak around 2025, and then continue to decrease.
Similarly, the daily electricity peak demands from now to 2032
are predicted by A bottom-up approach in EirGrid’s Ten-Year
Generation Capacity Statement\cite{GenerationCapacity}. It covers the forecasting values from now to 2032 in typical seasonal scenarios (winter, summer, and spring-autumn). Considering the increasing trend of annual power demand is steady, the Autoregressive Integrated Moving Average (ARIMA) is used to fill the missing values between 2033 and 2050. 15-minute-resolution electricity and gas demand curves are used for daily operational analysis. They are obtained from EirGrid’s smart grid dashboard and the Gas Network Ireland data dashboard in different seasonal scenarios \cite{SmartGridDashboard, GasconsumptionbyMarketSector}.
For the power transmission system, high voltage power lines over 110 kV are modelled, and the network topology, geographical locations, and electrical parameters (such as impedance) are obtained from EirGrid's Transmission Forecast Statement \cite{AllIslandTenYear}. The high-pressure gas transmission pipeline network (above 16 bar) is modelled, and similarly, the network topology, geographical locations, and pipeline length are obtained from Gas Network Ireland's Network Development Plan \cite{NetworkDevelopmentPlan}. Friction factors and diameters are assumed according to a dedicated case study of Ireland \cite{ekhtiari2022green}. The capacities of power lines are assumed to grow annually at the same rate as the demand growth.

\subsection{International green hydrogen trading model}
Besides consuming hydrogen/ammonia products by domestic energy systems, the surplus can be transported in the form of liquid by shipping to mainland Europe. Therefore, we propose an international hydrogen trading model, which utilises the supply-cost curves derived previously to simulate the hydrogen trading flows among the studied European countries from now to 2050. The supply potential and demands are important boundaries for determining the future landscape. On the supply side, for countries which have clearly outlined the scale of offshore wind farms at each or some critical years (such as Ireland \cite{FutureFrameworkforOffshore} and the UK \cite{Offshorewindnet}), these values are set as the maximum capacities for their offshore wind generation. For some countries which do not have fully outlined offshore wind capacities in their national policies in all critical years (such as the Netherlands and Spain), their tentative goals set in the European Union's non-binding agreement on offshore wind goals are used \cite{Offshorerenewableenergy}. 
The hydrogen and ammonia demands used in this study are predicted by the European Hydrogen Backbone Project in \cite{Analysingfuturedemand}. 

The objective of the proposed international green hydrogen trading model is to minimise the production and transportation costs while trying to meet all hydrogen/ammonia demands, as the details can be seen in Section 5 in Supplementary Information. The production cost is from the special-ordered set interpretation of supply-cost curves. The transportation cost includes annualised ship investment costs and operating fuel costs. The annualised ship cost determines the number of ships invested in each possible route, which further determines the maximum trading capacity among countries. 
The fuel cost is related to the length of the shipping route between the ports of exporting and importing countries from \cite{EMODnet}.
Due to the discontinuity of the hydrogen supply curves, the original problem is formulated as a mixed integer linear programming problem, and
is solved by the Gurobi commercial solver via MATLAB.

\backmatter

\section*{Acknowledgments}

This work was supported by the Newcastle University Academic Track (NUAcT) Fellowship.

\section*{Author contributions}

\section*{Competing interests}

The authors declare no competing interests.

\section*{Data availability}

The sources of the public meteorological dataset are provided in the References section. Data and methods for the model generation are provided in Supplementary Information.

\section*{Code availability}

Codes used in MATLAB for this study are available on Github \url{https://github.com/ShengWang-EE/HyExport}.

\begin{appendices}

\bmhead{Supplementary information}

\end{appendices}


\bibliography{sn-bibliography}


\end{document}